\documentclass[preprint,12pt,showkeys,nofootinbib,floatfix]{revtex4}
\usepackage{fancyhdr}
\usepackage{amssymb}
\usepackage{amsfonts}
\usepackage{amsmath}
\usepackage{graphicx}
\usepackage{subfigure}
\usepackage{tikz}
\usepackage[
            pdfstartview=FitH,
            bookmarksnumbered=true,
            bookmarksopen=true,
            colorlinks,
            pdfborder=001,
            linkcolor=blue,
            anchorcolor=blue,
            citecolor=red
            ]{hyperref}
\usepackage[toc,page,title,titletoc,header]{appendix}
\usepackage{graphics}
\usepackage{epsfig}
\usepackage{slashed}
\usepackage{latexsym}
\usepackage{syntonly}

\newcommand{\be}{\begin{equation}}
\newcommand{\ee}{\end{equation}}

\newcommand{\ran}{\rangle}
\newcommand{\Tr}{\mathrm{Tr}}
\begin{document}
\title{Information loss paradox revisited: farewell firewall?}
\author{Wen-Cong Gan$^{1}$}
\thanks{\text{Wen-cong\_Gan1@baylor.edu}}
\author{Fu-Wen Shu$^{2,3,1}$}
\thanks{Corresponding author; shufuwen@ncu.edu.cn}
%\thanks{E-mail address: shufuwen@ncu.edu.cn}
\affiliation{
$^{1}$ GCAP-CASPER, Physics Department, Baylor University, Waco, TX, 76798-7316, USA\\
$^{2}$Department of Physics, Nanchang University, No. 999 Xue Fu Avenue, Nanchang, 330031, China\\
$^{3}$Center for Relativistic Astrophysics and High Energy Physics, Nanchang University, No. 999 Xue Fu Avenue, Nanchang 330031, China}
\begin{abstract}
Unitary evolution makes pure state on one Cauchy surface evolve to pure state on another Cauchy surface. Outgoing Hawking radiation is only subsystem on the late Cauchy surface. The requirement that Hawking radiation to be pure amounts to require purity of subsystem when total system is pure. We will see this requirement will lead to firewall even in \textit{flat} spacetime, and thus is invalid. Information is either stored in the entanglement between field modes inside black hole and the outgoing modes or stored in correlation between geometry and Hawking radiation when singularity is resolved by quantum gravity effects. We will give a simple argument that even in semi-classical regime, information is (at least partly) stored in correlation between geometry and Hawking radiation.
\end{abstract}
\maketitle
\thispagestyle{fancy}        
\fancyhead{}                     
\lhead{Essay written for the Gravity Research Foundation 2020 Awards for Essays on Gravitation.}      
\chead{}
\rhead{}
\lfoot{}
\cfoot{\thepage}   %current page number
\rfoot{Submission date: March 31}
\renewcommand{\headrulewidth}{0pt}       
\renewcommand{\footrulewidth}{0pt}

\pagestyle{plain}
\cfoot{\thepage}

Firewall paradox is based on four assumptions \cite{Almheiri:2012rt}:
\begin{quote}

{Postulate 1:} {The process of formation and evaporation of a black hole, as viewed by a distant observer, can be described entirely within the context of standard quantum theory. In particular, there exists a unitary $S$-matrix which describes the evolution from infalling matter to outgoing Hawking-like radiation.}

{Postulate 2:} {Outside the stretched horizon of a massive black hole, physics can be described to good approximation by a set of semi-classical field equations.}

{Postulate 3:} To a distant observer, a black hole appears to be a quantum system with discrete energy levels. The dimension of the subspace of states describing a black hole of mass $M$ is the exponential of the Bekenstein entropy $S(M)$.

{Postulate 4:} A freely falling observer experiences nothing out of the ordinary when crossing the horizon.
\end{quote} The authors of \cite{Almheiri:2012rt} are referred to as AMPS. The first three postulates consist postulates of black hole complementary (BHC) \cite{Susskind:1993if}. The key reason that leads to Postulate 1 in \cite{Susskind:1993if}  is
\begin{quote}
If the black hole evaporates completely, that information would be lost, in violation of the rules of quantum theory.
\end{quote}

The Postulate 1 is equivalent to say Hawking radiation is pure if initially infalling matter is in pure state long before the black hole formed, because unitary evolution map pure state only to pure state. But distant observer can only access outgoing Hawking radiation. And outgoing Hawking radiation is only a subsystem because one ignores the part of the field modes inside black hole. This amounts to require subsystem to be pure in the background of the purity of the total system. We will see that it is exactly this requirement that leads to firewall paradox.

\section{Firewall in flat space-time}
\begin{figure}
\begin{center}
\includegraphics[height=2cm]{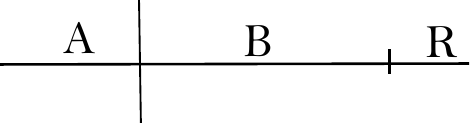}
\end{center}
\caption{Tripartite system $ABR$ in pure state $|vac\rangle$}\label{tripartite}
\end{figure} 

Let us consider a tripartite system $ABR$ in pure vacuum state $|vac\rangle$ in \textit {flat} spacetime (figure.(\ref{tripartite})). Then $A$ and $BR$ are highly entangled:
\be\label{smoothvac}
|vac\ran=\sum_n p_n|n\ran_A |n\ran_{BR},
\ee
Following a slightly different logic in \cite{Almheiri:2012rt}, if we assume subsystem $BR$ is also in pure state, i.e. $S_{BR}=0$, then by the Araki-Lieb triangle inequality
\begin{align}\label{Araki-Lieb}
S_{ABR}\le S_A+S_{BR},\\
S_{ABR}\ge |S_A-S_{BR}|,
\end{align}
we have $S_{ABR}=S_A$, then eq.(\ref{smoothvac}) requires 
\be
S_{ABR}=S_A=0.
\ee

So we must have 
\be
I_{A,BR}\equiv S_A+S_{BR}-S_{ABR}=0.
\ee 
The mutual information $I_{A,BR}$ is 0 if and only if $\rho_{ABR}=\rho_A\otimes \rho_{BR}$, which contradicts with the fact  that $A$ and $BR$ are highly entangled (cf. eq.(\ref{smoothvac})). And AMPS concludes that there is firewall between subsystem $A$ and $BR$ to break the entanglement. Follow this logic, we see that there exists firewall even in flat spacetime if we \textit{assume} the purity of subsystem $BR$ in the background of purity of the total system $ABR$. This strongly indicates the irrationality of the purity of subsystem $BR$, i.e. we cannot require subsystem to be pure when we have a pure total system. 

\begin{figure}
\begin{center}
\includegraphics[scale=0.2]{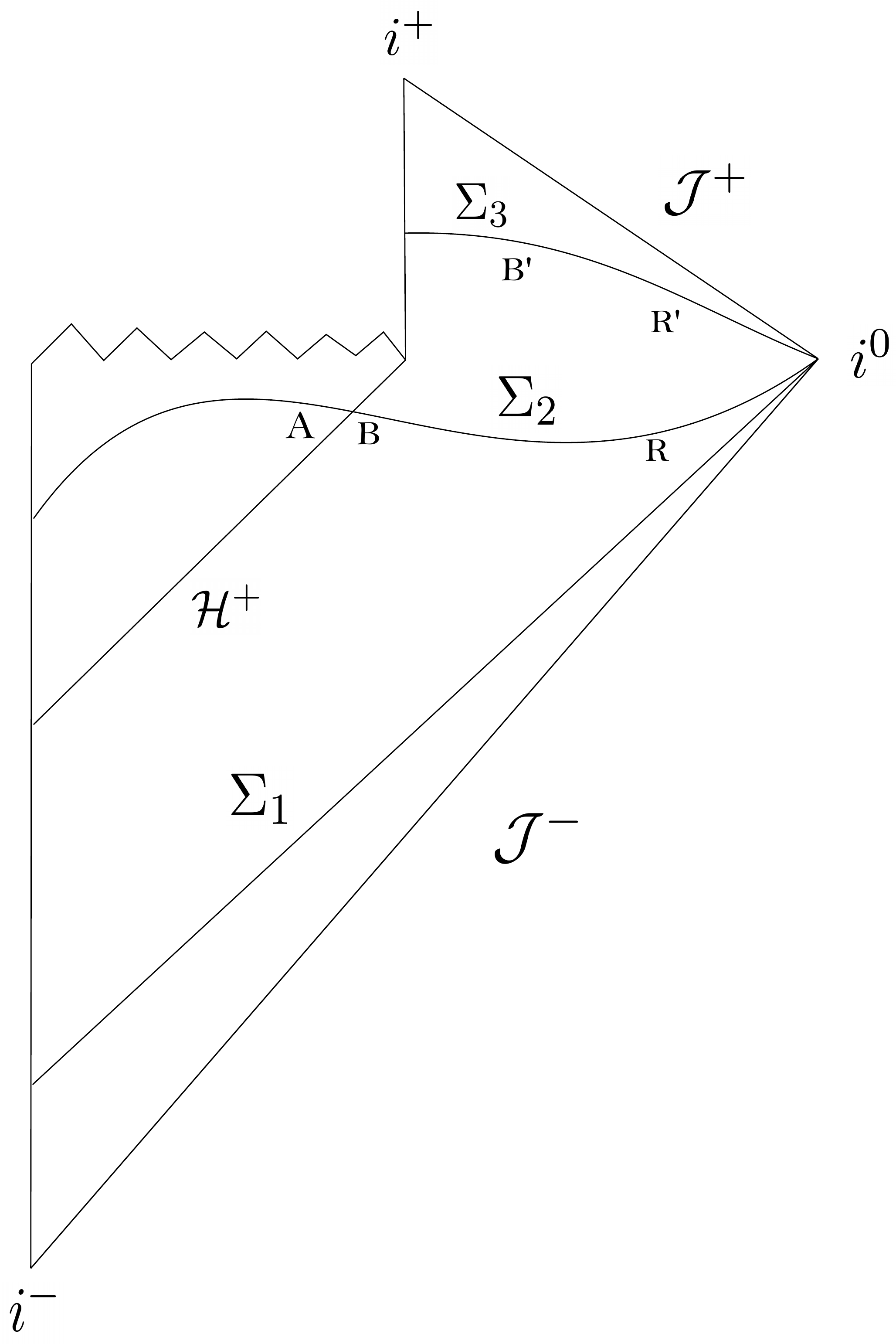}
\end{center}
\caption{$A$ represents the field modes inside horizon (including infalling Hawking quanta and infalling matter), $B$ is late outgoing Hawking quanta (after Page time), $R$ is early outgoing Hawking quanta.}\label{alice}
\end{figure} 

\section{Firewall in black hole space-time with singularity}
When applied to the argument of AMPS, $A$ represents the field modes inside horizon (including infalling Hawking quanta and infalling matter), $B$ is late outgoing Hawking quanta (after Page time \cite{Page:1993wv}), $R$ is early outgoing Hawking quanta, they are on late Cauchy surface $\Sigma_2$  (figure.(\ref{alice})). The infalling matter which forms black hole is pure on early Cauchy surface $\Sigma_1$. We quote another postulate from \cite{Unruh:2017uaw}, and call it Postulate 5:
\begin{quote}
In the case of quantum field theory, the full system consists of the quantum field observables over all of spacetime, or, equivalently—assuming deterministic evolution—the quantum field observables in a neighborhood of any Cauchy surface, $\Sigma$.
\end{quote}
Then unitary evolution makes state $\rho_{ABR}$ of total system on late Cauchy surface $\Sigma_2$ to be pure. But Postulation 1 in AMPS require subsystem $BR$ to be in pure state. If we assume Araki-Lieb triangle inequality holds also in the vicinity of black hole horizon (which is guaranteed by Postulate 2), then this leads to firewall. But as we pointed out in the previous section, we cannot require subsystem to be pure when we have a pure total system. We conclude that the assumption of the purity of Hawking radiation $BR$ is invalid, i.e. Hawking radiation is always mixed when we consider pure state of total system. Then where does the information go?

As stated in \cite{Unruh:2017uaw} (which is actually can be derived from Postulate 2 and Postulate 5):
\begin{quote}
The entanglement between the field in two such causally complementary regions always occurs in quantum field theory, no matter what the spacetime or the (physically acceptable) state.
\end{quote} 
When the total system is pure, the information is stored in the entanglement between subsystem $A$ and $BR$. Then $BR$ evolves to system $B'R'$ on non-Cauchy surface $\Sigma_3$. Even \textit{after} the black hole has evaporated totally, quantum fields on $\Sigma_3$ is still entangled with field modes inside the past black hole \cite{Unruh:2017uaw}. Thus the information on non-Cauchy surface $\Sigma_3$ alone cannot determine the initial state of black hole and information loss occurs on $\Sigma_3$ without violating fundamental physics laws \cite{Unruh:2017uaw}. Therefore there is no need to assume Postulate 1 as in \cite{Almheiri:2012rt,Susskind:1993if}.

Remark: Postulate 4 in \cite{Almheiri:2012rt} requires purity of subsystem $AB$ to make horizon smooth. However,  the smoothness of horizon can be also guaranteed by the only requirement of the purity of the total system $ABR$, as it implies the existence of entanglement between $A$ and $BR$. 

%In the previous paragraph, we argue that unitary requires purity of total system ABR, thus requires the existence of entanglement between A and BR, which can also guarantee the smoothness of horizon.

\section{Firewall in black hole space-time without singularity}

\begin{figure}
\begin{center}
\includegraphics[scale=0.2]{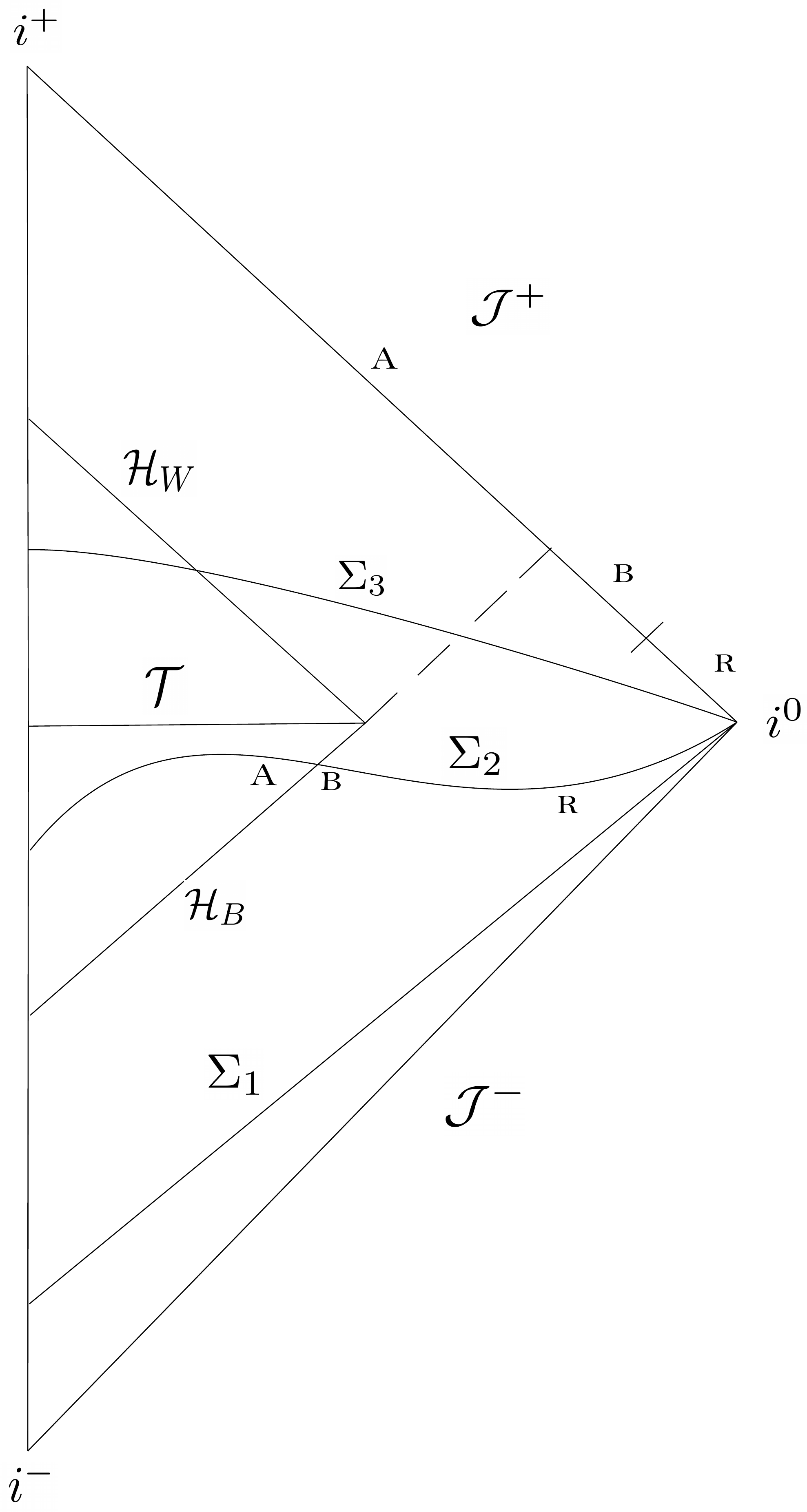}
\end{center}
\caption{An effective model describing black hole interior where singularity is replaced by a smooth transition surface $\mathcal T$, and it is connected with a white-hole-like future with asymptotically flat exterior.}\label{b2w}
\end{figure}

Recently, a quantum effective model describing black hole interior is proposed \cite{Haggard:2014rza,Ashtekar:2018cay,Martin-Dussaud:2019wqc}, in this model singularity is replaced by a smooth transition surface $\mathcal T$, and it is connected with a white-hole-like future with asymptotically flat exterior\footnote{For other discussions on that quantum gravity effects resolves black hole singularity see also \cite{Saini:2014qpa,Greenwood:2008ht,Bogojevic:1998ma,Wang:2009ay}.}  (figure.(\ref{b2w})). 

The causal past of future infinity $i^+ \bigcup \mathcal{J^+}$ is the whole space-time in figure (\ref{b2w}), thus we can safely admit Postulate 1 that asymptotic observer on future infinity $i^+ \bigcup \mathcal{J^+}$ sees the evolution to be unitary. This is in contrast with the case in figure (\ref{alice}), where the causal past of future infinity $i^+ \bigcup \mathcal{J^+}$ is \textit{not} the whole space-time and thus there is no reason to admit Postulate 1 in this kind of space-time, because asymptotic observer on future infinity $i^+ \bigcup \mathcal{J^+}$ can only access \textit{part} of the information of the whole space-time. Even we admit Postulate 1 in the case of figure (\ref{b2w}), the unitary makes the whole system $ABR$ on the future infinity $i^+ \bigcup \mathcal{J^+}$ to be pure, not $BR$ itself. Thus in this case there is still no conflict with the Araki-Lieb triangle inequality, so no firewall will appear. In this case, information is stored in entanglement between subsystem $A$ and $BR$ even after white hole has exploded.

Remark: We have seen that the causal structure is very important. Depending on whether the causal past of future infinity $i^+ \bigcup \mathcal{J^+}$ is the whole space-time or \textit{not}, different causal structure of the space-time will make Postulate 1 valid or not. Both cases, however, exclude BR as a pure state, thus exclude firewall as well.

\section{Entanglement with geometry}
Back-reaction of Hawking radiation is always ignored since it is small. Until Parikh and Wilczek \cite{Parikh:1999mf} use tunneling to explain dynamical mechanism of Hawking radiation, then people realize the importance of back-reaction of Hawking radiation on the geometry to deal with information paradox.

Recently, authors in \cite{Perez:2014xca} pointed out that based on Loop Quantum Gravity (LQG) the information also can be stored in the correlation between the Hawking quanta and the fundamental pre-geometric structures\footnote{For other discussions on correlation between geometry and field modes based on non-LQG method see also \cite{Hutchinson:2013kka,Saini:2015dea,Acquaviva:2017xqi}.} \cite{Perez:2014xca}. And for course-grained observer who cannot access fundamental degrees of freedom, he/she will also see mixed state of Hawking quanta. In this case, there is also no need for firewall to make Hawking radiation pure.

One might argue that the argument in \cite{Perez:2014xca} is based on quantum gravity effect which occurs only in Plank regime near the transition surface \cite{Ashtekar:2018cay} which replaced the singularity in quantum gravity or in Plank regime at the end of evaporation where semi-classical approximation breaks down. How to solve the firewall paradox in semi-classical regime? We will give an argument that even in semi-classical regime information will be (at least partly) stored in correlation between Hawking quanta and geometry.

In figure (\ref{b2w}), long before the black hole forms, the spacetime is Minkowski and the ``in" state of the field modes we set to be vacuum state and thus pure. Taking into account the state of geometry, we choose the state on the Cauchy surface $\Sigma_1$ to be $\rho_{FG}=|vac\rangle \langle vac| \otimes |Minkowski\rangle \langle Minkowski|$, where $F$ represents field modes and $G$ represents geometry. Then we have 
\begin{align}
\rho_F&=\Tr_{G}(\rho_{FG})=|vac\rangle \langle vac|,\label{density}\\
\rho_G&=\Tr_{F}(\rho_{FG})=|Minkowski\rangle \langle Minkowski|,
\end{align}
and we choose the state on the Cauchy surface $\Sigma_2$ to be some unknown state $\rho_{F'G'}$ with the properties:
\begin{align}
\rho_{F'}&=\Tr_{G'}(\rho_{F'G'})=\rho_{Hawking},\\
\rho_{G'}&=\Tr_{F'}(\rho_{F'G'})=\rho_{black  \ hole}.
\end{align}

If the evolution is unitary 
\be
\rho_{F'G'}=U\rho_{FG}U^{\dagger},
\ee
then entropy $S=-\Tr (\rho \ln \rho)$ is invariant:
\be
S_{FG}=S_{F'G'}.
\ee
And the mutual information is increasing:
\be
I_{F'G'}-I_{FG}=(S_{F'}+S_{G'}-S_{F'G'})-(S_{F}+S_{G}-S_{FG})> 0,\label{mutual}
\ee
because $S_{F'}\ge S_{F}$ and $S_{G'}> S_{G}$.
Thus even though there may be no entanglement between field modes and geometry at the beginning, there will be such entanglement after black hole has formed. And information will be stored in entanglement between field modes and geometry even in semi-classical regime. Thus we cannot assume Hawking quanta to be pure and there is no need for firewall to make it pure.

Remark: even though there are some comments \cite{Bouhmadi-Lopez:2019hpp} on the asymptotic behavior of the quantum effective model in \cite{Ashtekar:2018cay}, the derivation in \cite{Perez:2014xca} and our argument based on eq.(\ref{density}) to eq.(\ref{mutual}) do not rely on asymptotic behavior in the model.

\section{Conclusion}
In summary, if we insist on Postulate 5, i.e. the unitary from Cauchy surface to Cauchy surface, then we cannot assume Hawking radiation to be pure and thus there is no need of firewall. Information is either stored in the entanglement between field modes inside black hole and the outgoing modes even after black hole has totally evaporated or stored in correlation between geometry and Hawking radiation when singularity is resolved by quantum gravity effects.

\section*{Acknowledgements}
We thank Anzhong Wang for valuable discussions. W.-C.G. is supported by Baylor University through the Baylor Physics graduate program. This work was supported in part by the National Natural Science Foundation of China under grant numbers 11975116, and Jiangxi Science Foundation for Distinguished Young Scientists under grant number 20192BCB23007.
\\
\ \\


\begin{thebibliography}{99}
%\cite{Almheiri:2012rt}
\bibitem{Almheiri:2012rt} 
  A.~Almheiri, D.~Marolf, J.~Polchinski and J.~Sully,
  ``Black Holes: Complementarity or Firewalls?,''
  JHEP {\bf 1302}, 062 (2013)
  doi:10.1007/JHEP02(2013)062
  [arXiv:1207.3123 [hep-th]].
  %%CITATION = doi:10.1007/JHEP02(2013)062;%%
  %1004 citations counted in INSPIRE as of 02 Mar 2020

%\cite{Susskind:1993if}
\bibitem{Susskind:1993if} 
  L.~Susskind, L.~Thorlacius and J.~Uglum,
  ``The Stretched horizon and black hole complementarity,''
  Phys.\ Rev.\ D {\bf 48}, 3743 (1993)
  doi:10.1103/PhysRevD.48.3743
  [hep-th/9306069].
  %%CITATION = doi:10.1103/PhysRevD.48.3743;%%
  %888 citations counted in INSPIRE as of 25 Mar 2020
  
  %\cite{Page:1993wv}
\bibitem{Page:1993wv} 
  D.~N.~Page,
  ``Information in black hole radiation,''
  Phys.\ Rev.\ Lett.\  {\bf 71}, 3743 (1993)
  doi:10.1103/PhysRevLett.71.3743
  [hep-th/9306083].
  %%CITATION = doi:10.1103/PhysRevLett.71.3743;%%
  %375 citations counted in INSPIRE as of 28 Mar 2020
  
%\cite{Unruh:2017uaw}
\bibitem{Unruh:2017uaw} 
  W.~G.~Unruh and R.~M.~Wald,
  ``Information Loss,''
  Rept.\ Prog.\ Phys.\  {\bf 80}, no. 9, 092002 (2017)
  doi:10.1088/1361-6633/aa778e
  [arXiv:1703.02140 [hep-th]].
  %%CITATION = doi:10.1088/1361-6633/aa778e;%%
  %91 citations counted in INSPIRE as of 02 Mar 2020


  
%\cite{Haggard:2014rza}
\bibitem{Haggard:2014rza} 
  H.~M.~Haggard and C.~Rovelli,
  ``Quantum-gravity effects outside the horizon spark black to white hole tunneling,''
  Phys.\ Rev.\ D {\bf 92}, no. 10, 104020 (2015)
  doi:10.1103/PhysRevD.92.104020
  [arXiv:1407.0989 [gr-qc]].
  %%CITATION = doi:10.1103/PhysRevD.92.104020;%%
  %118 citations counted in INSPIRE as of 03 Mar 2020

%\cite{Ashtekar:2018cay}
\bibitem{Ashtekar:2018cay} 
  A.~Ashtekar, J.~Olmedo and P.~Singh,
  ``Quantum extension of the Kruskal spacetime,''
  Phys.\ Rev.\ D {\bf 98}, no. 12, 126003 (2018)
  doi:10.1103/PhysRevD.98.126003
  [arXiv:1806.02406 [gr-qc]].
  %%CITATION = doi:10.1103/PhysRevD.98.126003;%%
  %33 citations counted in INSPIRE as of 02 Mar 2020

%\cite{Martin-Dussaud:2019wqc}
\bibitem{Martin-Dussaud:2019wqc} 
  P.~Martin-Dussaud and C.~Rovelli,
  ``Evaporating black-to-white hole,''
  Class.\ Quant.\ Grav.\  {\bf 36}, no. 24, 245002 (2019)
  doi:10.1088/1361-6382/ab5097
  [arXiv:1905.07251 [gr-qc]].
  %%CITATION = doi:10.1088/1361-6382/ab5097;%%
  %6 citations counted in INSPIRE as of 03 Mar 2020

%\cite{Saini:2014qpa}
\bibitem{Saini:2014qpa}
A.~Saini and D.~Stojkovic,
``Nonlocal (but also nonsingular) physics at the last stages of gravitational collapse,''
Phys. Rev. D \textbf{89}, no.4, 044003 (2014)
doi:10.1103/PhysRevD.89.044003
[arXiv:1401.6182 [gr-qc]].
%22 citations counted in INSPIRE as of 09 Aug 2020

%\cite{Greenwood:2008ht}
\bibitem{Greenwood:2008ht}
E.~Greenwood and D.~Stojkovic,
``Quantum gravitational collapse: Non-singularity and non-locality,''
JHEP \textbf{06}, 042 (2008)
doi:10.1088/1126-6708/2008/06/042
[arXiv:0802.4087 [gr-qc]].
%34 citations counted in INSPIRE as of 09 Aug 2020

%\cite{Bogojevic:1998ma}
\bibitem{Bogojevic:1998ma}
A.~Bogojevic and D.~Stojkovic,
``A Nonsingular black hole,''
Phys. Rev. D \textbf{61}, 084011 (2000)
doi:10.1103/PhysRevD.61.084011
[arXiv:gr-qc/9804070 [gr-qc]].
%27 citations counted in INSPIRE as of 09 Aug 2020

%\cite{Wang:2009ay}
\bibitem{Wang:2009ay}
J.~E.~Wang, E.~Greenwood and D.~Stojkovic,
``Schrodinger formalism, black hole horizons and singularity behavior,''
Phys. Rev. D \textbf{80}, 124027 (2009)
doi:10.1103/PhysRevD.80.124027
[arXiv:0906.3250 [hep-th]].
%30 citations counted in INSPIRE as of 09 Aug 2020

%\cite{Parikh:1999mf}
\bibitem{Parikh:1999mf} 
  M.~K.~Parikh and F.~Wilczek,
  ``Hawking radiation as tunneling,''
  Phys.\ Rev.\ Lett.\  {\bf 85}, 5042 (2000)
  doi:10.1103/PhysRevLett.85.5042
  [hep-th/9907001].
  %%CITATION = doi:10.1103/PhysRevLett.85.5042;%%
  %1322 citations counted in INSPIRE as of 25 Mar 2020

%\cite{Hawking:1974sw}
%\bibitem{Hawking:1974sw} 
%  S.~W.~Hawking,
%  ``Particle Creation by Black Holes,''
%  Commun.\ Math.\ Phys.\  {\bf 43}, 199 (1975)
%  Erratum: [Commun.\ Math.\ Phys.\  {\bf 46}, 206 (1976)].
%  doi:10.1007/BF02345020, 10.1007/BF01608497
%  %%CITATION = doi:10.1007/BF02345020, 10.1007/BF01608497;%%
%  %8115 citations counted in INSPIRE as of 02 Mar 2020

%\cite{Perez:2014xca}
\bibitem{Perez:2014xca} 
  A.~Perez,
  ``No firewalls in quantum gravity: the role of discreteness of quantum geometry in resolving the information loss paradox,''
  Class.\ Quant.\ Grav.\  {\bf 32}, no. 8, 084001 (2015)
  doi:10.1088/0264-9381/32/8/084001
  [arXiv:1410.7062 [gr-qc]].
  %%CITATION = doi:10.1088/0264-9381/32/8/084001;%%
  %29 citations counted in INSPIRE as of 03 Mar 2020
  
  %\cite{Hutchinson:2013kka}
\bibitem{Hutchinson:2013kka}
J.~Hutchinson and D.~Stojkovic,
``Icezones instead of firewalls: extended entanglement beyond the event horizon and unitary evaporation of a black hole,''
Class. Quant. Grav. \textbf{33}, no.13, 135006 (2016)
doi:10.1088/0264-9381/33/13/135006
[arXiv:1307.5861 [hep-th]].
%23 citations counted in INSPIRE as of 19 May 2020

%\cite{Saini:2015dea}
\bibitem{Saini:2015dea}
A.~Saini and D.~Stojkovic,
``Radiation from a collapsing object is manifestly unitary,''
Phys. Rev. Lett. \textbf{114}, no.11, 111301 (2015)
doi:10.1103/PhysRevLett.114.111301
[arXiv:1503.01487 [gr-qc]].
%44 citations counted in INSPIRE as of 19 May 2020

%\cite{Acquaviva:2017xqi}
\bibitem{Acquaviva:2017xqi}
G.~Acquaviva, A.~Iorio and M.~Scholtz,
``On the implications of the Bekenstein bound for black hole evaporation,''
Annals Phys. \textbf{387}, 317-333 (2017)
doi:10.1016/j.aop.2017.10.018
[arXiv:1704.00345 [gr-qc]].
%6 citations counted in INSPIRE as of 19 May 2020

%\cite{Bouhmadi-Lopez:2019hpp}
\bibitem{Bouhmadi-Lopez:2019hpp} 
  M.~Bouhmadi-López, S.~Brahma, C.~Y.~Chen, P.~Chen and D.~h.~Yeom,
  ``Comment on "Quantum Transfiguration of Kruskal Black Holes",''
  arXiv:1902.07874 [gr-qc].
  %%CITATION = ARXIV:1902.07874;%%
  %9 citations counted in INSPIRE as of 03 Mar 2020


\end{thebibliography}
\end{document}